# Modeling indoor-level non-pharmaceutical interventions during the COVID-19 pandemic: a pedestrian dynamics-based microscopic simulation approach

Yao Xiao, xiaoyao9@sysu.edu.cn

School of Intelligent System Engineering, Sun Yat-Sen University, Shenzhen, Guangdong, China

Mofeng Yang, mofeng@umd.edu

Maryland Transportation Institute, Department of Civil and Environmental Engineering, University of Maryland at College Park, Maryland, USA

Zheng Zhu*, zhuzheng@ust.hk

Department of Civil and Environmental Engineering, Hong Kong University of Science and Technology, Hong Kong, China

Hai Yang, cehyang@ust.hk

Department of Civil and Environmental Engineering, Hong Kong University of Science and Technology, Hong Kong, China

Lei Zhang, lei@umd.edu

Maryland Transportation Institute, Department of Civil and Environmental Engineering, University of Maryland at College Park, Maryland, USA

Sepehr Ghader, sghader@umd.edu

Maryland Transportation Institute, Department of Civil and Environmental Engineering, University of Maryland at College Park, Maryland, USA

*Corresponding Author, email: zhuzheng@ust.hk*

**ABSTRACT**

Mathematical modeling of epidemic spreading has been widely adopted to estimate the threats of epidemic diseases (i.e., the COVID-19 pandemic) as well as to evaluate epidemic control interventions. The indoor place is considered to be a significant epidemic spreading risk origin, but existing widely-used epidemic spreading models are usually limited for indoor places since the dynamic physical distance changes between people are ignored, and the empirical features of the essential and non-essential travel are not differentiated. In this paper, we introduce a pedestrian-based epidemic spreading model that is capable of modeling indoor transmission risks of diseases during people's social activities. Taking advantage of the before-and-after mobility data from the University of Maryland COVID-19 Impact Analysis Platform, it's found that people tend to spend more time in grocery stores once their travel frequencies are restricted to a low level. In other words, an increase in dwell time could balance the decrease in travel frequencies and satisfy people's demand. Based on the pedestrian-based model and the empirical evidence, combined non-pharmaceutical interventions from different operational levels are evaluated. Numerical simulations show that restrictions on people's travel frequency and open-hours of indoor places may not be universally effective in reducing average infection risks for each pedestrian who visit the place. Entry limitations can be a widely effective alternative, whereas the decision-maker needs to balance the decrease in risky contacts and the increase in queue length outside the place that may impede people from fulfilling their travel needs.

*Key words: epidemic spreading, pedestrian dynamics, travel demand*

## 1. Introduction

Epidemic diseases are public enemies of human beings. With the capability of transmission among the human population, epidemic diseases have taken many more lives than all armed conflicts during the last century (Adda, 2016). Although modern medical science has significantly increased the survival chance of an individual patient, the epidemics of stubborn viruses, such as SARS, H1N1, and the newly emerging COVID-19, still raise great public concerns. For instance, the fast-global spread of COVID-19 in the year 2020 drags down medical systems in many countries by quickly depleting limited medical resources (i.e., beds, medical staff, and medicine), and thereby many patients have to rely on self-immunity. Moreover, the pandemic and the accompanying measures (e.g. lock-down policy) also raise negative influences on the productivity, cause significant social costs (Almond, 2006), and even shock the global financial market and social economy[1]. Consequently, it is of great importance to block or at least slow down the spread of epidemics.

The protracted war between humans and diseases can involve multiple disciplines, such as medical science, social science, and transportation science. This is because the transmission of an epidemic disease among the human population has a complex mechanism related to various factors, including the contagiousness of the virus, physical contacts between individuals, the medical or social prevention strategies, and so on. In the era of social networks and globalization, the control and prevention of epidemics can be ever more challenging. Namely, advanced transportation infrastructure and convenient mobility services can send a person to anywhere within the human society, leading to fast city-to-city and even country-to-country disease transmission. Diverse social activities (i.e., from community activities, such as parties, to city-wide activities, such as working and shopping, then to international activities, such as the Olympics) cause more frequent mass gatherings, and thereby increase the chance of physical contacts between infected individuals and susceptible individuals. Recent empirical studies have illustrated the positive relationship between epidemic spreading and modern transportation, such as airlines (Olsen et al., 2003) and railways (Adda, 2016), as well as the breakout of epidemics due to social activities (Alzeer, 2009; Khan et al., 2010).

---

[1] Reported by the New York Times, COIVD-19 is responsible for stocks plunge all around the world in 2020 https://www.nytimes.com/2020/03/05/business/stock-market-covid-19.html.

Given the complex transmission process and the importance of epidemic control and prevention, researchers have developed mathematical epidemiology to depict the spread of epidemics through a plethora of factors. At the macroscopic level, researchers focus on the infection and recovery process of disease among the population. Aggregate social and medical-related factors such as population infection and recovery rate, long-distance travel rate, and age distribution are usually used for building mathematical formulations to model epidemic spreading dynamics (Earn, 2008). Although the macroscopic models established the research discipline of mathematical epidemiology, it may ignore details in modeling the social interactions among individuals. For instance, the models cannot describe the spreading dynamics within or between small communities. Therefore, macroscopic approaches can be insensitive in evaluating epidemic control strategies, or they may require strong assumptions to overcome the incapability. The limitations were partially addressed by activity-based models that estimate individual-level temporal-spatial travels related to social activities in a city or a small community (Eubank et al., 2004). The existing models are helpful but still have two shortcomings. Firstly, they are imperfect to describe the epidemic spreading process with time-varying individual physical distances during pedestrian movements. With cities getting increasingly crowded, the mass gathering can exist everywhere and raise heated concerns for epidemic spreading (Khan et al., 2012). In this manner, the capability of modeling pedestrian-level epidemic spreading is indispensable. Secondly, the empirical features of essential and non-essential travel are usually not discussed and differentiated. Even though non-pharmaceutical interventions can restrict the number of social activities, they could lead to side-effects such as increasing average indoor dwell times as well as exposure time to infected pedestrians, especially for the essential travel demand. To this end, the features of essential travel demand should be incorporated in the design of non-pharmaceutical interventions.

Nowadays, with the fast development of computational technologies and individual-level location sensors, pedestrian dynamics are more preferable in describing individual decisions and actions in mass gathering scenarios. By integrating pedestrian dynamics with mathematical epidemiology, researchers can model the spread of epidemics in certain places at the microscopic level, such as a railway station and a restaurant. Thereby, a variety of non-pharmaceutical interventions related to crowed movements can be evaluated, such as restricting the open-hours, restricting the number of entries into places, or dividing and redirecting of the pedestrian flow. In this paper, we introduce a pedestrian-based epidemic spreading model to ascertain the transmission risk of epidemics in a general indoor place and evaluate the performance of combined non-pharmaceutical interventions from different operational levels (i.e., population-level and indoor-level) according to the practical travel demand. With realistic before-and-after mobility data from the University of Maryland (UMD) COVID-19 Impact Analysis Platform (IAP), the travel pattern changes during the pandemic can be analyzed using people's number of visits, duration of the visits, and arrival times to indoor places.

The remainder of this paper is organized as follows. Section 2 provides a literature review on the state-of-the-art macroscopic and microscopic (i.e., activity-based and pedestrian-based) approaches in mathematical epidemiology. Section 3 introduces the pedestrian-based epidemic spreading model adopted in this paper. In Section 4, we analyze people's travel pattern changes before and during the COVID-19 pandemic based on the UMD COVID-19 IAP. With the pedestrian-based epidemic spreading model and the empirical analysis, we evaluate the indoor transmission risk for both baseline scenarios and the scenarios with combined non-pharmaceutical interventions. Finally, conclusions and future work are drawn in Section 6.

## 2. Literature review

Mathematical epidemiology modeling started in the 1900s when Hamer (1906) introduced a discrete-time model to analyze the recurrence of measles epidemics. Later on, Ross (1910) developed differential equation models for the prediction and control of malaria. Inspired by early work, the well-known susceptible–infected–recovered (SIR) model was developed (Kermack and McKendrick, 1927). The SIR model provides a foundation for aggregate mathematical epidemiology. The basic idea is to simplify the complex disease spreading process by the relationship among a few aggregate factors: the number of people who are susceptible to be infected $S(t)$, the number of infectors who spreads the disease $I(t)$, the number of recovery people who will not be infected again $R(t)$, the transmission rate $\beta$, and the recovery rate $\alpha$, wherein $t$ denotes time (Kermack and McKendrick, 1927). Nonlinear ordinary differential equations are used to depict the SIR dynamics of epidemic diseases in the population[2]. The SIR model was extended to various differential equation-based models, including the susceptible–infected–recovered–susceptible (SIRS) model (Hethcote, 1976), the susceptible–exposed–infected–recovered (SEIR) model (Liu et al., 1987), the immunity–susceptible–exposed–infected–recovered (MSEIR) model (Schuette, 2003), structural population-based SIR model (Cross et al., 2007), and the stochastic SEIR model (Black and McKane, 2010). Based on key threshold quantities, such as basic reproduction ratio (usually denoted as $R_0$) (Diekmann et al., 1990) and contact number (Hethcote, 1976), researchers attempted to obtain the stationary condition in the epidemic spreading process. The aggregate models contribute to the design and analysis of macroscopic disease control strategies. However, one comment limitation is that they extremely simplify the disease transmission process (Hethcote, 2009).

Since the spreading of an epidemic is largely related to people's social activities, individual-level disease transmission study has become a key component in mathematical epidemiology. Microscopic modeling of the disease spreading process began with the improvement of computational power and simulation-based technologies. Kleczkowsi and Grenfell (1999) simulated the socio-geographical spread of diseases based on the cellular automata approach. By assuming that each person has a small social-network, Eubank et al. (2004) adopted bipartite graphs to model people's contact patterns in the city and analyzed the spread and control of smallpox. The authors generated population and the social network based on socioeconomic data and shown the advantages of microscopic simulations in reproducing real-world scenarios. Several microscopic epidemic spreading models were developed, such as cell-based models (Karl et al., 2014), social-network-based models (Milne et al., 2008), and space-time activity-based models (Yang and Atkinson, 2008). These models simulate contacts between individuals during social activities and estimate the transmissions based on predefined parameters such as infectious distance and transmission rate.

---

[2] The early aggregate model utilizes three differential equations: $\frac{dS(t)}{dt} = -\beta S(t)I(t)$, $\frac{dI(t)}{dt} = \beta S(t)I(t) - \alpha I(t)$, and $\frac{dR(t)}{dt} = \beta I(t)$ (Kermack and McKendrick, 1927).

It is undeniable that these macroscopic and microscopic models are capable of evaluating medium-to-large scale epidemic control strategies, including school closure (Milne et al., 2008), household quarantine (Longini et al., 2005), case isolation (Ferguson et al., 2006), antiviral prophylaxis (Rizzo et al., 2008), travel restrictions (Epstein et al., 2007), vaccination (Carrat et al., 2006), and their combinations (Nuno et al., 2007). Although the results indicated that the combination of activity restriction strategies is significantly effective in delaying the peaks and reducing the number of total infected people, they might be less sound in two ways. First, the aforementioned studies ignored the differences between people's non-essential travel demand and essential travel demand for economic and living activities, such that people need to work and shop to maintain basic living requirements. Second, they may not be sensitive to the effect of people's physical movements on disease transmission in specific indoor places, such as in a hospital, a restaurant, and a cruise.

Pedestrian-based epidemic spreading models, which examine person-to-person disease transmission dynamics leveraging pedestrian crowd movements, are well fit for the aforementioned limitations of macroscopic and activity-based microscopic models. With the capability of modeling time-varying movements of individuals, pedestrian dynamics have been utilized in various disciplines, such as walking facility design (Helbing et al., 2005, Qu et al. 2014), hub network optimization (Gao et al., 2014; Hanseler et al., 2020), mass activity organization (Hoogendoorn et al., 2004; Alnaulsi et al. 2014), and crowd emergency evacuation (Helbing et al., 2000; Zheng et al., 2017). Pedestrian dynamics started to attract attention in the field of mathematical epidemiology from the last decade. It has been shown that the simulation of crowd pedestrian movements can result in the same aggregate epidemic spreading patterns compared with macroscopic models (Johansson et al., 2012). To the best of our knowledge, the investigations of epidemic spreading from the microscopic pedestrian movement perspective are insufficient, and most of them focused on some specific situations, such as air travel (Namilae et al., 2017) and cruise (Fang et al., 2020). The integration between pedestrian dynamics and epidemiology is still getting heated, and it will refine our understanding of the spread and control of epidemics (Johansson and Goscè, 2014). In this paper, we attempt to provide perspectives on pedestrian-based epidemic spreading models by simulating various combined non-pharmaceutical interventions in a general indoor place.

## 3. Pedestrian-based epidemic spreading model

The proposed pedestrian-based microscopic model calculates the infection probabilities based on a person-to-person disease transmission mechanism. There are two parts in the microscopic model, which are respectively the pedestrian movement module and the disease transmission risk module. The former module reproduces the general pedestrian crowd movement in public indoor places, while the latter module describes the disease transmission risk from an infector to the susceptible individuals. We provide detailed formulas for the two modules in this section.

### 3.1 Pedestrian dynamics

The social force model (Helbing et al., 1995, 2000; Gayathri et al., 2017) is one of the most popular methods in reproducing pedestrian crowd movements, and it has been successfully applied as the fundamental model of microscopic pedestrian crowd simulations in commercial software, such as PTV Vissim and MassMotion[3]. The social force model is applied for the simulation of pedestrian dynamics, in which a pedestrian is regarded as an object affected by several forces from goal, neighbors, and obstacles,

$$\boldsymbol{F}_i(t) = \boldsymbol{F}_i^{drv}(t) + \sum_j \boldsymbol{F}_{i,j}^{ped}(t) + \sum_w \boldsymbol{F}_{i,w}^{obs}(t) \quad (1)$$

where $\boldsymbol{F}_i(t)$ is a two-dimensional vector that represents the resultant of all the external forces of pedestrian $i$ at time $t$. For the remaining of this paper, we adopt bold notations for vectors. The three components of $\boldsymbol{F}_i(t)$ are explained as follows.

Firstly, $\boldsymbol{F}_i^{drv}(t)$ indicates the attraction force to the goal at time $t$, and it reflects the desire of pedestrian $i$ (with a mass of $m_i$) to maintain a certain walking speed $v_i^0$ towards certain direction $\boldsymbol{e}_i^0$ in a relaxation time $\tau_i$. Let $v_i(t)$ and $\boldsymbol{e}_i(t)$ represent the instant speed and unit direction vector respectively, the formulation of $\boldsymbol{F}_i^{drv}(t)$ is given by:

$$\boldsymbol{F}_i^{drv}(t) = m_i \frac{v_i^0 \boldsymbol{e}_i^0 - v_i(t)\boldsymbol{e}_i(t)}{\tau_i} \quad (2)$$

Here $v_i^0$ and $v_i(t)$ are non-negative real numbers, and $\boldsymbol{e}_i^0$ and $\boldsymbol{e}_i(t)$ are two-dimensional unit direction vectors which can be represented by $(\cos\theta, \sin\theta), \theta \in U(0, 2\pi)$.

Secondly, $\boldsymbol{F}_{i,j}^{ped}(t)$ indicates the interaction force between the objective pedestrian $i$ and the neighboring pedestrian $j$. Generally, it is a repulsive force and comes in two parts, the social force on the psychological level and the contact force on the physical level (shown in Eq. (3)).

---

[3] See websites https://www.ptvgroup.com/en/solutions/products/ptv-vissim/ and https://www.oasys-software.com/products/pedestrian-simulation/massmotion/ for more details about PTV VISSIM and MassMotion.

$$\boldsymbol{F}_{i,j}^{ped}(t) = \left\{A \exp\left(\frac{r_i + r_j - d_{i,j}(t)}{B}\right) + Cg\left(r_i + r_j - d_{i,j}(t)\right)\right\}\boldsymbol{n}_{i,j}(t) \quad (3)$$

where $r_i$, $r_j$ and $d_{i,j}(t) = ||\boldsymbol{l}_i(t) - \boldsymbol{l}_j(t)||$ represent the radius of pedestrian $i, j$ and the instant distance between the centers of both pedestrians ($\boldsymbol{l}_i(t)$ and $\boldsymbol{l}_j(t)$ denotes the instant location, i.e., coordinates, of the two pedestrians). $A$, $B$ and $C$ are parameters, and $A$ is 2000 newtons, $B$ equals 0.08 meters, and $C$ equals 120000 kilogram per second squared, $\boldsymbol{n}_{i,j}(t)$ denotes the unit direction vector from pedestrian $j$ to pedestrian $i$ at time $t$. And $g(x)$ is a function that does not equal zero only if the pedestrians are touched,

$$g(x) = \begin{cases} x, & \text{if } x > 0 \\ 0, & \text{otherwise} \end{cases} \quad (4)$$

Thirdly, $\boldsymbol{F}_{i,w}^{obs}(t)$ indicates the instant interaction force between the objective pedestrian $i$ and the wall/obstacle $w$, as presented in Eq. (5).

$$\boldsymbol{F}_{i,w}^{obs}(t) = \left\{A \exp\left(\frac{r_i - d_{i,w}(t)}{B}\right) + Cg\left(r_i - d_{i,w}(t)\right)\right\}\boldsymbol{n}_{i,w}(t), \quad (5)$$

where, similar to Eq. (3), $d_{i,w}(t)$ represents the instant distance to the obstacle $w$, and $\boldsymbol{n}_{i,w}(t)$ denotes the instant direction from the wall/obstacle $w$ to the pedestrian $i$.

Accordingly, the motion of the pedestrian is driven under Newton's second law with a second-order dynamics function (Moussaid et al. 2011), and the velocity of pedestrian $i$ is determined as,

$$\frac{dv_i(t)\boldsymbol{e}_i(t)}{dt} = \frac{\boldsymbol{F}_i(t)}{m_i} \quad (6)$$

and the location $\boldsymbol{l}_i(t)$ of pedestrian $i$ is determined as

$$\frac{d\boldsymbol{l}_i(t)}{dt} = v_i(t)\boldsymbol{e}_i(t) \quad (7)$$

Furthermore, a random walking process is formulated through the adjustment of the original desired direction $\boldsymbol{e}_i^0$ to simulate the crowd movements in a space with boundaries (i.e., indoor place). In our model, the pedestrian is needed to change the original desired direction when he/she is closely approaching the boundary. Suppose one pedestrian with a desired direction $\boldsymbol{e}_i^0 = (cos\theta, sin\theta)$ is going to hit the boundary, and the unit normal direction vector of the boundary is $\boldsymbol{e}_i^b = (cos\theta^b, sin\theta^b)$. Then the model will force him/her to change the direction, and the new desired direction is arbitrarily obtained as,

$$\boldsymbol{e}_i^0 \in \{(\cos\theta', \sin\theta')|\theta' \in U(0, 2\pi), \cos\theta'\cos\theta^b + \sin\theta'\sin\theta^b < 0\} \quad (8)$$

Note that the new desired direction of pedestrian $i$ is required to be any direction that is away from hitting the boundary.

### 3.2 The risk of epidemic spreading

In this subsection, the risk of epidemic spreading is calculated based on a microscopic person-to-person transmission mechanism. According to the current investigation on COVID-19 (World Health Organization, 2020), the route of person-to-person transmission of this disease is via respiratory droplets; and close contacts with someone who has sneezing, coughing, or other respiratory symptoms can put a person on a high risk of being infected. WHO suggested that anyone who has face-to-face contacts with a probable or confirmed case within 1 meter for over 15 minutes, or who has direct physical contacts with a probable or confirmed case shall be identified as a close contact. Also, the Centers for Disease Control and Prevention (CDC) believes that maintaining good social distance (6 feet, which is about 1.8 meters) is important in preventing the spread of COVID-19. Consequently, the exposure time in the infection area is regarded as a core index to indicate the transmission risk of epidemics.

In our model, both getting close to infected individuals and locating in the coughing area would cause potential risk on epidemic spreading, and the general exposed time $T_i$ of pedestrian $i$ is calculated by accumulating the contact exposed time $T_i^{ct}$ and the cough exposed $T_i^{cg}$, such that

$$T_i = T_i^{ct} + T_i^{cg} \quad (9)$$

The first term, i.e., contact exposed time $T_i^{ct}$, it is calculated by directly summing up the physical contacts between the target pedestrian and every infected individual in the place (shown in Eq. (10)).

$$T_i^{ct} = \sum_{t=t_i^{enter}}^{t_i^{exit}} \sum_{m=1}^{M(t)} \lambda_{i,m}(t) \quad (10)$$

where $t_i^{enter}$ and $t_i^{exit}$ respectively denotes the place enter time and exit time of pedestrian $i$, and $M(t)$ indicates the number of infected pedestrians in the place at time $t$. $\lambda_{i,m}(t)$ indicates the status of whether pedestrian $i$ is exposed in the influence area of infected pedestrian $m$, and it is obtained as,

$$\lambda_{i,m}(t) = \begin{cases} 1, & \text{if } d_{i,m}(t) \leq D_c - r_m \\ 0, & \text{otherwise} \end{cases} \quad (11)$$

where $D_c$ indicates the cut-off distance of the infector influence area, and $d_{i,m}(t)$ represents the instant distance between the centers of pedestrian $i$ and infected pedestrian $m$.

Compared with the contact infection, the transmission through coughing usually has a great influence area but a shorter existence time. Suppose the influence area of coughing is a circle area with the center to be the coughing individual, the second term, i.e., cough exposed time $T_i^{cg}$, it is calculated as follows,

$$T_i^{cg} = \sum_{t=t_i^{enter}}^{t_i^{exit}} \sum_{n=1}^{N(t)} \mu_{i,n}(t) \qquad (12)$$

where $N(t)$ indicates the number of existing coughing areas (i.e., number of infected individuals who are coughing) in the place at time $t$, and $\mu_{i,n}(t)$ indicates the status whether pedestrian $i$ is exposed in the influence area of coughing area $n$, and it is obtained as,

$$\mu_{i,n}(t) = \begin{cases} 1, & \text{if } d_{i,n}(t) \leq R_n \\ 0, & \text{otherwise} \end{cases} \qquad (13)$$

where $d_{i,n}(t)$ indicates the distance from the pedestrian $i$ to the center of coughing area $n$, and $R_n$ is the radius of the coughing area (i.e., also referred to as mean coughing distance).

## 4. Travel demand analysis based on UMD COVID-19 IAP

The research team utilizes mobility data from the UMD COVID-19 IAP[4] (Zhang et al., 2020(a)). The platform aggregates mobile device location data from more than 100 million anonymized monthly active users across the United States on a daily basis. Figure 1 illustrates the observed mobility changes during the COVID-19 pandemic[5] using UMD COVID-19 IAP. Before the National Emergency declaration on Marth 13th, the number of trips per person was stable and high; while, after the declaration, the number of trips per person first dramatically dropped (March 13th - March 22nd, 2020) and then maintained at a low level (after March 22nd).

---

[4] See our website https://data.covid.umd.edu for more details.
[5] Detailed methodology related to data processing and plotting can be found in Zhang et al. (2020(b)). Moreover, The U.S. government proclaimed a national state of emergency on March 13th, 2020. And by April 10th, only 8 states had not issued "Stay-at-home" orders (Xiong et al. 2020).

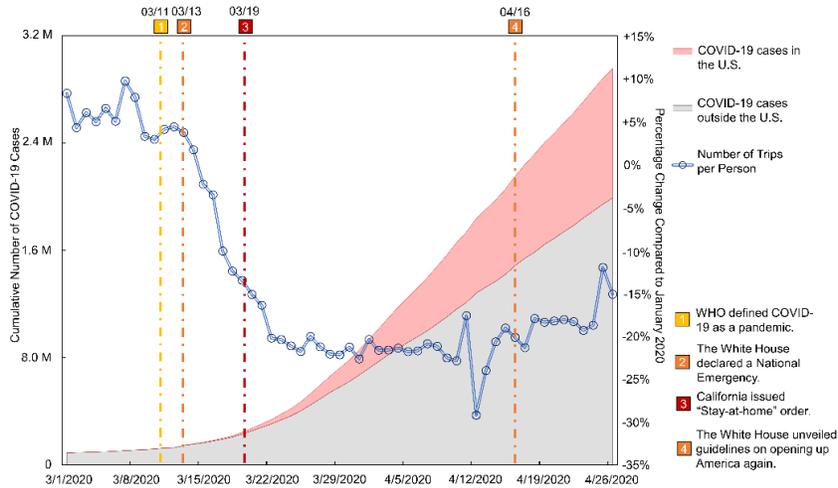

Figure 1. Observed mobility changes using UMD COVID-19 IAP

Based on 56 days of mobile device location data (March 1st - April 26th, 2020), we analyze the travel demand with respect to daily mean dwell time and the number of visits for grocery shopping in Figure 2[6]. The daily data points illustrate notable distinctive travel patterns, which can be grouped into two clusters. The first cluster (blue points) refers to the mobility data before March 22nd, about three days after the state of California issued the "stay-at-home" order. In this cluster, the number of daily visits has a negligible impact on the mean dwell time. While, in the second cluster (red points, refers to the mobility data after March 22nd), we find a significant negative relationship between dwell time and the number of visits. Two regression curves for the two clusters are given in Figure 2, and comparing a non-clustering regression with the clustering-based regression, the R-Squared value increases from 0.124 to 0.527. The clustered travel pattern indicates insightful travel behavior findings, which are summarized below.

- The number of visits notably drops as people receive the government's call and perceive the high transmission risk of COVID-19 during their travel activities.
- For grocery shopping trips, since the two travel patterns can be classified mainly by the number of visits, there exists a specific visiting number that is regarded as the watershed of travel demand.
- Once the number of visits is above the watershed, the grocery visits consist of both essential shopping activities for basic living requirements and non-essential shopping activities, such that restrictions on travel frequency would not hurt people's essential travel demand.
- Once the number of visits is below the watershed, the grocery visits only consist of essential shopping activities, and as the government further restrict people's travel activities, visitors tend to shop longer to fulfill their living requirements.

---

[6] The location data is classified by the attribute of the place, such as grocery stores, pharmacy stores, restaurants, etc. We focus on grocery shopping because it is closely related to people's essential travel demand.

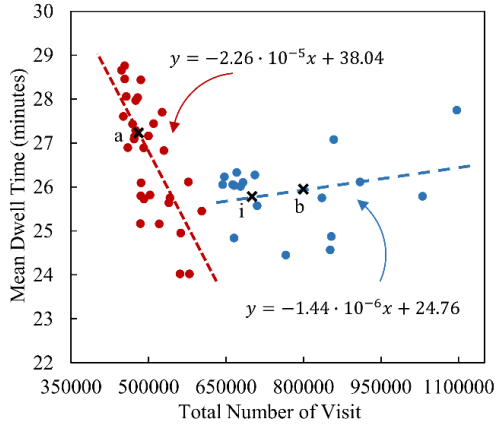

Figure 2. Daily mean dwell time versus number of visits

Besides the "stay-at-home" order, most grocery stores adopt an early closure policy. For instance, Walmart U.S. stores adjusted its operating hours to 7:00 to 20:30 during the COVID-19 pandemic[7]. Based on the aforementioned mobility data, we summarize the distributions of customer arrival times for days before many stores implemented early closure policies (i.e., Match 24th is selected as the watershed date) and days after early closure policies. The before-and-after customer arrival patterns are depicted in Figure 3, in which we observe notable shifts of the peak period beginning time from around 16:00 (red) to around 13:00 (blue). Consequently, we claim that an early closure will lead to earlier and wider peak periods, which is referred to as a peak spreading phenomenon.

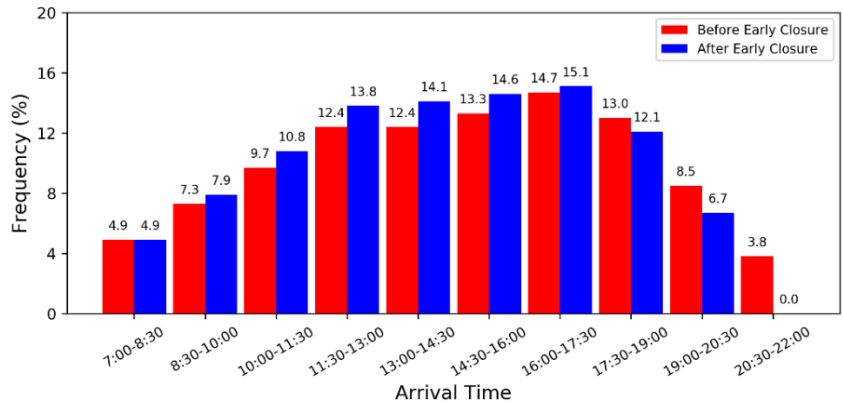

Figure 3. Arrival time distribution

The conclusions in the travel demand analysis provide us a solid behavior foundation in designing non-pharmaceutical interventions. That is, if the government would like to further restrict people's travel, it should expect an increase in average dwell time and the underlying transmission risk for the visitors who have to conduct an essential trip. And store managers need to concern the peak spreading phenomenon while implementing an open-hours restriction policy.

---

[7] More information can be found at https://repstevenreick.com/2020/03/20/covid-19-grocery-store-hours/.

## 5. Estimating indoor transmission risks of COVID-19

Plenty of approaches have been successfully applied in simulating the infectious disease transmission through the macroscopic perspectives and activity-based microscopic perspectives. However, to the best of our knowledge, very few approaches have been used to investigate the performance of different non-pharmaceutical interventions in general indoor places. A critical reason shall be that people seldom meet disease that has both a long period of incubation and strong infectious ability, like the emerging COVID-19. Due to the worldwide pandemic of COVID-19, lock-down policies are adopted by numerous countries and the residents are suggested to stay at home as long as possible. These policies would lead to a sharp descend of mass gathering in public places, and the disease spreading by contact or droplet spreading should also reduce. Nevertheless, it cannot be ignored that there is still a demand for the essential travels even in the pandemic, e.g., buying necessities and seeking medical treatment. Besides, the coronavirus can exist for quite a long time among human beings considering the overall trend of the pandemic fact, and more demands for public places would emerge not only from the perspective of the economy but also from that of human nature. Therefore, understand the transmission risk in public indoor places and figure out what is the optimal non-pharmaceutical interventions are quite significant.

In this section, a general indoor place is adopted to study the transmission risk and potential influence of non-pharmaceutical interventions based on crowd dynamics simulation. Section 5.1 describes the basic simulation settings and the performance measures of indoor transmission risk. Section 5.2 evaluates the transmission risks of the real-world scenarios obtained from the UMD COVID-19 IAP, which are referred to as baseline scenarios. In Section 5.3, we estimate the performance of combined non-pharmaceutical interventions from different operational levels and investigate the optimal strategy that could further slow down the spread of COVID-19.

### 5.1 Basic simulation settings

The practical place settings vary according to different places, and one specific setting might affect virus transmission risks. For instance, a restaurant has sitting tables, and a shopping center has shop units (absorbing spots). However, focusing on the specific setting is not necessary for this paper. This is because it is unlikely to include all types of indoor geometries in the study, and it is less significant than some other parameters, e.g., area and the spatial distribution of pedestrians. Here, we introduce a public indoor place with a $40 \times 30$ squared meter room but no obstacles inside, and the pedestrians would freely walk inside the place. Each pedestrian is represented by a circle of 0.2 meters (i.e., $r_i = 0.1$ meters,) and the relaxation time $\tau$ and the desired speed $v_i^0$ are 0.5 seconds and 1.34 meter per second, respectively[8]. The infection rate of the residents is set to be 0.5% in the simulation, i.e., 0.5% of the pedestrians are infected with COVID-19. It is supposed that these people will infect others through contact and cough approaches. For the contact approach, the cut-off distance for the exposure $D_c$ is 1 meter according to the suggestion of WHO (World Health Organization, 2020). For the cough approach, an infected pedestrian is supposed to have a cough every 15 seconds. By referring to the mean coughing distance among the volunteers in (Loh et al., 2020), $R_n$ is set to be 2.5 meters, and the infectious time of coughing averagely lasts for 15 seconds and 5 seconds in indoor places and out-of-door areas, respectively. The exposure time of each people inside the contact and the cough infection area can be calculated via Eqs. (9) – (13).

---
[8] It is noticed that the practical walking speeds for different places can be unequal. For instance, people usually sit at a fixed position in the restaurant, but when it is a supermarket or a stadium they will generally keep walking or even running. However, it is found that the speed in simulation is not a decisive factor for the total infection time (see appendix A, and total infection time is defined in Eq. (14)).

The epidemic spreading process is illustrated in Figure 4. The red, orange, and green dots represent the infected people, the exposed susceptible people, and the unexposed health people, respectively. The yellow area indicates the influence area of both contact and coughing.

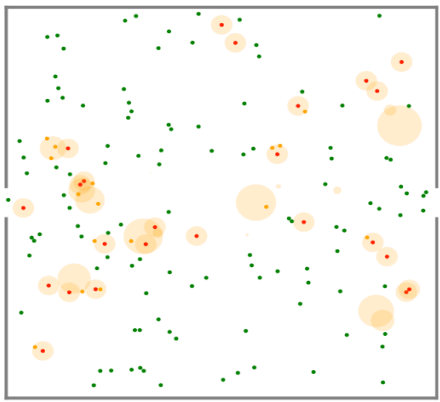

Figure 4. Sketch map of pedestrian-based epidemic spreading process.

To quantify the indoor transmission risk of epidemics, the exposed time of a susceptible individual (i.e., $T_i$ defined in Eq. (9)) is calculated. Two critical measurements are calculated based on $T_i$: the total infection time and the number of high-risk exposed people. The total infection time $T^{inf}$ is the summation of exposed time for all the susceptible individuals who have visited in the general indoor place.

$$T^{inf} = \sum_i T_i \qquad (14)$$

The number of high-risk exposed people $C^{risk}$ refers to the number of individuals whose exposure time is over $\bar{T}$ during the visit.

$$C^{risk} = \sum_i \phi(T_i), \qquad (15)$$

where $\phi(T_i)$ is calculated as,

$$\phi(T_i) = \begin{cases} 1, & \text{if } T_i > \bar{T} \\ 0, & \text{otherwise} \end{cases}, \tag{16}$$

and we assume $\bar{T}$ to be 20 seconds in this paper[9].

## 5.2 Evaluation of baseline scenarios

Based on the travel demand analysis in Section 4, we consider three baseline scenarios as real-world benchmarks and estimate the corresponding indoor transmission risks. Firstly, we utilize a "before" scenario to represent the situation before the US government announced the national emergency. From the mobility data, we select a general day representing scenario "before" and assume 8000 visitors are coming to the public place during the open-hours (i.e., 7:00 to 22:00); the dwell time of the visitors is around 25.9 minutes (point 'b' in Figure 2). Secondly, an "intermediate" scenario is adopted to represent the situation when people have reduced non-essential trips to some extent, but the total number of visits is still above the watershed of people's travel demand. For scenario "intermediate", we assume that 7000 pedestrians visit the place and their dwell time is 25.8 minutes (point 'i' in Figure 2). In both scenarios "before" and "intermediate", we assume the early closure policy is not implemented and the arrival times of pedestrians follow the red bar histogram in Figure 3. Thirdly, we use an "after" scenario to represent the real-world condition with restrictive self-shielding and early closure policies, and people's essential demand is hurt. The total number of visits and the average dwell time is 4800 and 27.4 minutes respectively, which comes from the medium condition (i.e., point 'a') of the red cluster in Figure 2[10]. Due to the implementation of the early closure policy (i.e., from 7:00 to 20:30), the arrival times will follow the blue bar histogram in Figure 3. Note that if the arrival times of some pedestrians are nearby the closing time, we allow them to finish their planned dwell time.

Figure 5 depicts the indoor transmission risks under the three baseline scenarios. The total infection time ($T^{inf}$, red bars) and number of high-risk exposed pedestrians ($C^{risk}$, blue bars) are shown in Figure 5(a), while the average infection time ($\frac{T^{inf}}{I}$, red bars) and the high-risk ratio ($\frac{C^{risk}}{I} \times 100\%$, blue bars) are illustrated in Figure 5(b), wherein $I$ denotes the number of visits. Comparing the scenarios, we find that both the total and the pedestrian average transmission risk will significantly decrease as people reduce their travel frequencies (i.e., measured by the number of visits). Although the dwell time increase from scenario "intermediate" to scenario "after", the average risk for an individual pedestrian in the place still becomes lower because of a huge reduction of visits (i.e., by around 31.4% from 7000 to 4800).

---

[9] An exposed time of over 20 seconds is risky because the infected individual may have infective actions (e.g., coughs and talks) by chance and transmit virus to the susceptible individual.

[10] Note that the exact values of the data points in Figure 2 stand for the aggregate data of numerous grocery stores, which might not reflect the number of visits for one indoor place. To make a general case study, we assume a constant scaler of 0.01, such that the point with 800000 visits in Figure 2 represents the "before" scenario with 8000 visits to the general indoor place. Similarly, the "intermediate" and "after" scenarios are denoted by points with 700000 and 480000 visits in Figure 2, respectively.

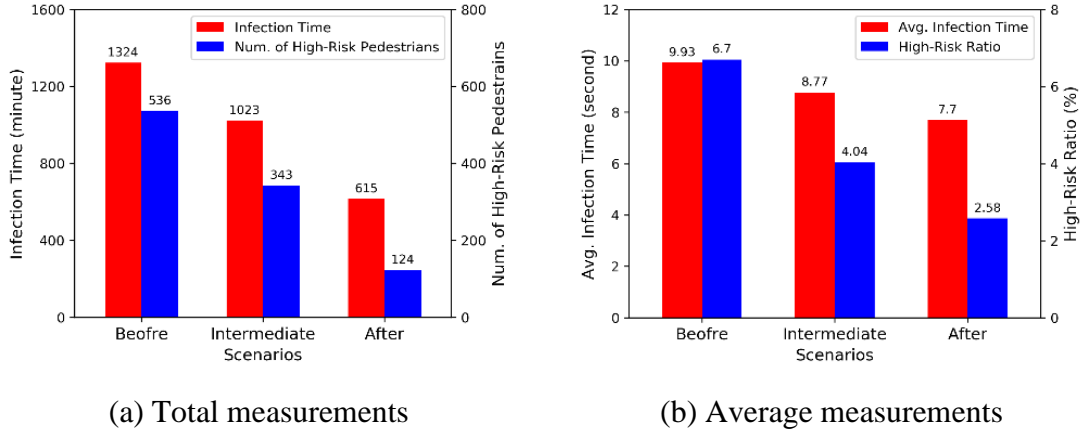

(a) Total measurements  (b) Average measurements

Figure 5. Results of the baseline scenarios

The results of the baseline scenarios imply the effect of activity restrictions on epidemic control. Although the total indoor infection risk largely decreases, the average infection risk is still undesirable (the average infection time is still 7.7 seconds, and the high-risk ratio is 2.58%). It's can't be ignored that visitors would come from different households, and fewer members from the same household would visit grocery stores to reduce the risk of infections during the activity restrictions. However, according to the current results on the average infection risk, this could still lead to infections of visitors from various households as well as serious transmissions within the infected households and small communities. Since this paper only cares about the indoor transmission process of diseases (i.e., COVID-19), the infection risks at household and community levels are out of scope. We attempt to seek good-intended combined interventions that can furtherly slow down the indoor spreading of diseases with a small high-risk ratio.

**5.3 Evaluation of combined non-pharmaceutical interventions**

Regarding the baseline scenario "after" as the current real-world situation, we explore possible pros and cons of implementing combined non-pharmaceutical interventions from different operational levels. Three levels of interventions are proposed and the meanings and specific parameter settings are given below:

**Self-shielding policy (population-level).** People are required by the government to reduce travel frequency during the pandemic. As the number of visits is below the watershed (i.e., Figure 2), the dwell time of visitors will correspondingly increase to fulfill the essential demand. Three strategies are respectively adopted: "S0" is with 6000 visitors and an average dwell time of 24.7 minutes, "S1" is with 4800 visitors and a 27.4 minutes' dwell time, and "S2" has 3600 visitors whose dwell time is 30.0 minutes. The dwell times are obtained based on the fitted linear relationship of the red points (i.e., essential travel demand) in Figure 2.

**Open-hours restriction policy (indoor-level).** An early closure policy is implemented by the manager of the indoor place, and it's found that earlier and wider peak periods can be formed according to the practical results in Figure 3. Here, two operational strategies are considered for this policy: "O0" is without early closure and the distribution of arrival times follow the red bars as shown in Figure 3, and the distribution of restricting strategy is implemented in "O1" such that the arrival times follow the blue bars.

**Entry limitation policy (indoor-level).** The maximum number of visitors inside the indoor place is limited by the manager to reduce the epidemic spreading risk among crowds. The extra visitors have to wait in a queue outside the indoor place until some other visitors leave the place, and they would maintain a physical distance of over 6 feet in the queue as suggested by CDC. We consider two strategies: "E0" is without entry limitation, and in "E1" the manager allows a maximum number of 150 visitors (each visitor averagely owns 8 m$^2$ in the general indoor place) in the place simultaneously. At the closing time, visitors who are still waiting in the queue are forced to cancel the activity.

The combined interventions are denoted by a sequence of intervention codes in the "S-O-E" format. For instance, scenario "S0-O1-E0" represents the combination of strategies "S0", "O1", and "E0" for the three operational levels. Note that the scenario "S1-O1-E0" is identical to the baseline scenario "after". Figure 6 depicts the performance of all 12 possible combinations of non-pharmaceutical interventions with respect to the total measurements $T^{inf}$ and $C^{risk}$ (i.e., Figure 6(a)) and the average measurements $\frac{T^{inf}}{I}$ and $\frac{C^{risk}}{I} \times 100\%$ (i.e., Figure 6(b)).

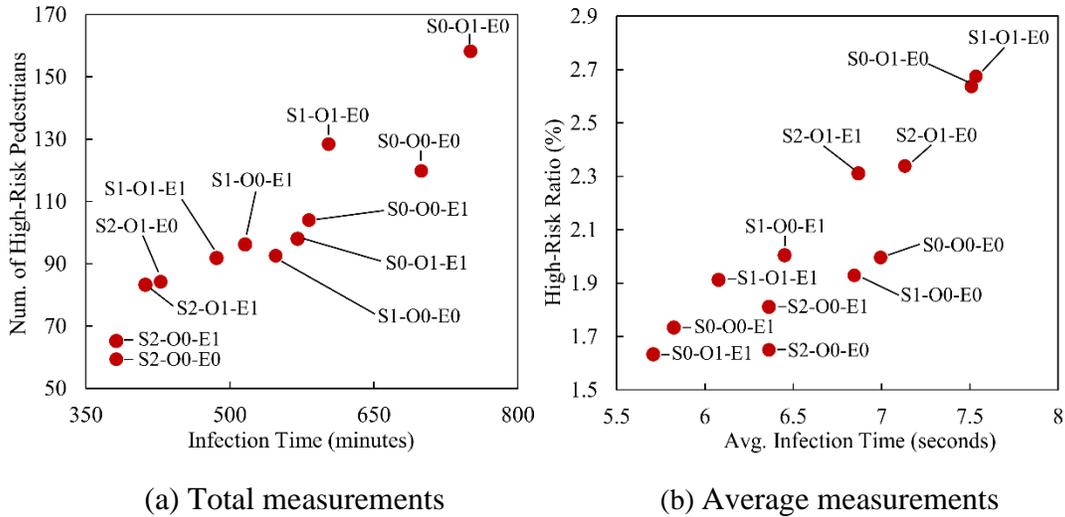

(a) Total measurements  (b) Average measurements

Figure 6. Results of the combined non-pharmaceutical interventions

Through a parallel comparison of the scenarios "S0", "S1" and "S2" which own different number of visitors, it can be found that the total infection risk (i.e., infection time and the number of high-risk pedestrians) monotonically fall with the decrease of visitor numbers, whereas the trend is not that significant for the average infection risk (i.e., average infection time and high-risk ratio). For instance, the average infection risk in "S1-O1-E0" is greater than that of "S0-O1-E0". In other words, increasing the magnitude of the self-shielding policy (i.e., further reducing people's travel frequency) might not be effective in decreasing individual infection risk. For the essential trips, the reduction in the number of visits will lead to higher dwell times, and the positive and negative effects (i.e., positive effects from fewer visits and negative effects from longer dwell times) on individual infection risk might cancel out.

Within each self-shielding level (i.e., "S0", "S1", and "S2"), the restriction on open-hours (i.e., "O1") leads to the increase of transmission risk. A possible reason is considered to be that the gathering of people during peak hours in the indoor places leads to the growing of pedestrian density (i.e., the inverse of person-to-person distance) in the indoor place, and the phenomenon just enhance the chance of exposure to infected pedestrians. Thereby, the combination "O1-E0" leads to the highest transmission risk for both the total and the average measurements.

Regarding the entry limitation policy (i.e., "E1"), it can notably slow down the spreading of COVID-19 during grocery shopping trips with self-shielding policies "S0" and "S1" as the number of visits is at a medium-to-high level. However, the queueing process might last until the closing time, so that some visitors cannot enter the place to fulfill their travel needs. Namely, 12.5%, 21.3%, 4.8%, and 13.9% of the visitors are still queued outside the place in combinations "S0-O0-E1", "S0-O1-E1", "S1-O0-E1" and "S1-O1-E1", respectively. Once the number of pedestrians is low (i.e., "S2"), the effectiveness of entry limitation is marginal (i.e., comparing "S2-O0-E0" with "S2-O0-E1", and "S2-O1-E0" with "S2-O1-E1").

Given "S1-O1-E0" as the benchmark, decision-makers can further restrict people's travel frequency or implement an entry limitation policy to achieve a more reliable control of epidemics. The results of this paper raise a missing concern in epidemic control. That is, due to people's essential travel demand, travel restrictions may not be universally effective in controlling the spreading of epidemics, especially on individual infection probabilities. An early closure might increase the indoor transmission risk, which is owing to the peak spreading of pedestrian arrivals. Although the open-hours restriction policy might not be effective for epidemic control in this paper, it could be a necessary concern for the place manager to keep the employees from the disease. In cases with a medium-to-high number of visits, entry limiting is remarkably effective, regardless of people's essential demand and the potential risk outside the place. While decision-makers may need to concern the length of the queue such that the last queued people can enter the place before its closing time. To satisfy all the stakeholders (i.e., pedestrians, the manager of the indoor place, and the government officials), good coordination is needed during the implementation of combined interventions.

## 6. Conclusions and Discussion

Microscopic mathematical epidemiology is becoming heatedly preferable due to its capability of modeling community-level epidemic spread and control interventions. Most existing microscopic models estimate the disease transmission process through the simulation of individual social activities, but generally ignore the dynamics in physical contacts due to pedestrian movements. Additionally, current research efforts on epidemic control might dismiss the distinction between people's essential and non-essential travel demand during the implementation of activities restrictions. In this paper, we attempt to address these research issues by utilizing a pedestrian-based epidemic spreading model coupled with real-world travel data during the COVID-19 pandemic. The pedestrian-based model provides a mathematical basis for estimating disease transmission risks for indoor activities, and the dataset shows the impacts of activity restriction interventions on people's travel patterns concerning their essential travel demand. Based on the proposed model and the empirical behavioral findings, the transmission risk of COVID-19 during indoor activities for reality (i.e., before, intermediate, and after the "stay-at-home" call) and scenarios with combined non-pharmaceutical interventions (i.e., self-shielding, open-hours restriction, and entry limitation) can be evaluated.

Based on the before-and-after travel demand analysis, we find that people's grocery shopping activities consist of both essential and non-essential trips. When the government begins to implement a self-shielding policy, people first tend to reduce non-essential trips and the average dwell time basically remains unchanged. However, once the number of visits is below some watershed, people will only conduct essential grocery shopping trips; and a smaller number of visits will lead to a longer dwell time to satisfy visitors' essential demand (i.e., food purchase). Besides, an early closure policy could change the distribution of arrivals and bring a peak spreading phenomenon of arrivals.

The results of the pedestrian-based indoor COVID-19 transmission simulations are insightful. Firstly, because of the essential travel demand feature, a self-shielding policy may not bring down individual infection risks. Secondly, from a system-level epidemic control perspective, an open-hours restriction policy could increase the indoor transmission risk of COVID-19 due to the aforementioned peak spreading phenomenon. Thirdly, an entry limitation policy is effective when the number of visits is large. However, the manager of the indoor place needs to take a balance between the number of people waiting outside and the potential decrease in transmission risks. The balance is important because people may need to maintain basic living requirements by entering the place. In addition, it's found that the performance of combined non-pharmaceutical interventions can be improved via good coordination among the decision-makers.

There are still some limitations to the current modeling approach, which can be fulfilled in the future. Firstly, the modeling about the pedestrian-level epidemic transmission process is still not comprehensive. For instance, we haven't considered the heterogeneity of individuals in terms of their walking speeds and preferred social distances. Secondly, the general place can be extended to various specific places, such as shopping malls, hospitals, and railway stations. That is, we can model a railway station by restricting people's movement directions, or a shopping center by adding absorbing pots as shops. Thirdly, the proposed model is currently not appropriate for the scenarios with different scales, while the activity-based or macroscopic models can be further integrated to fulfill the disease transmission dynamics in all scales.

**Acknowledgment**

The authors would like to thank and acknowledge our partners and data sources in this effort: (1) various mobile device location data providers; (2) Amazon Web Service and its Senior Solutions Architect, Jianjun Xu, for providing cloud computing and technical support; (3) computational algorithms developed and validated in a previous USDOT Federal Highway Administration's Exploratory Advanced Research Program project; and (4) COVID-19 confirmed case data from the Johns Hopkins University Github repository and sociodemographic data from the U.S. Census Bureau.

**References**

Adda, J. (2016). Economic activity and the spread of viral diseases: Evidence from high frequency data. *The Quarterly Journal of Economics*, *131*(2), 891-941.

Alnabulsi, H., Drury, J., (2014). Social identification moderates the effect of crowd density on safety at the Hajj. Proceedings of the National Academy of Sciences 111(25), 9091-9096.

Almond, D. (2006). Is the 1918 influenza pandemic over? Long-term effects of in utero influenza exposure in the post-1940 US population. *Journal of political Economy*, *114*(4), 672-712.

Alzeer, A.H. (2009). Respiratory tract infection during Hajj. *Annals of thoracic medicine*, *4*(2), 50.

Black, A.J., McKane, A. J. (2010). Stochasticity in staged models of epidemics: quantifying the dynamics of whooping cough. *Journal of the royal society interface*, *7*(49), 1219-1227.

Carrat, F., Luong, J., Lao, H., Sallé, A.V., Lajaunie, C., Wackernagel, H. (2006). A'small-world-like'model for comparing interventions aimed at preventing and controlling influenza pandemics. *BMC medicine*, *4*(1), 26.

Cross, P.C., Johnson, P.L., Lloyd-Smith, J.O., Getz, W.M. (2007). Utility of R 0 as a predictor of disease invasion in structured populations. *Journal of the Royal Society Interface*, *4*(13), 315-324.

Diekmann, O., Heesterbeek, J.A.P., Metz, J.A. (1990). On the definition and the computation of the basic reproduction ratio R 0 in models for infectious diseases in heterogeneous populations. *Journal of mathematical biology*, *28*(4), 365-382.

Earn, D.J. (2008). A light introduction to modelling recurrent epidemics. In *Mathematical epidemiology* (pp. 3-17). Springer, Berlin, Heidelberg.

Epstein, J.M., Goedecke, D.M., Yu, F., Morris, R.J., Wagener, D. K., Bobashev, G. V. (2007). Controlling pandemic flu: the value of international air travel restrictions. *PloS one*, *2*(5).

Eubank, S., Guclu, H., Kumar, V.A., Marathe, M.V., Srinivasan, A., Toroczkai, Z., Wang, N. (2004). Modelling disease outbreaks in realistic urban social networks. *Nature*, *429*(6988), 180-184.

Fang, Z., Huang, Z., Li, X., Zhang, J., Lv, W., Zhuang, L., ..., Huang, N. (2020). How many infections of COVID-19 there will be in the" Diamond Princess"-Predicted by a virus transmission model based on the simulation of crowd flow. *arXiv preprint arXiv:2002.10616*.

Ferguson, N.M., Cummings, D.A., Fraser, C., Cajka, J.C., Cooley, P.C., Burke, D.S. (2006). Strategies for mitigating an influenza pandemic. *Nature*, *442*(7101), 448-452.

Gao, Z.Y., Qu, Y.C., Li, X.G., Long, J.C., Huang, H.J. (2014). Simulating the Dynamic Escape Process in Large Public Places. Oper Res 62(6), 1344-1357.

Gayathri, H., Aparna, P.M., Verma, A. (2017). A review of studies on understanding crowd dynamics in the context of crowd safety in mass religious gatherings. *International journal of disaster risk reduction*, 25, 82-91.

Hamer, W.H. (1906). *Epidemic disease in England: the evidence of variability and of persistency of type*. Bedford Press.

Hänseler, F.S., van den Heuvel, J.P.A., Cats, O., Daamen, W., Hoogendoorn, S.P. (2020). A passenger-pedestrian model to assess platform and train usage from automated data. Transportation Research Part A: Policy and Practice 132(948-968.

Helbing, D., Buzna, L., Johansson, A., Werner, T. (2005). Self-organized pedestrian crowd dynamics: Experiments, simulations, and design solutions. Transportation Science 39(1), 1-24.

Helbing, D., Farkas, I., Vicsek, T. (2000). Simulating dynamical features of escape panic. *Nature,* 407(6803), 487-490.

Helbing, D., Molnar, P. (1995). Social force model for pedestrian dynamics. *Phys. Rev. E,* 51(5), 4282-4286.

Hethcote, H.W. (1976). Qualitative analyses of communicable disease models. *Mathematical Biosciences*, *28*(3-4), 335-356.

Hethcote, H.W. (2009). The basic epidemiology models: models, expressions for R0, parameter estimation, and applications. In *Mathematical understanding of infectious disease dynamics* (pp. 1-61).

Hoogendoorn, S.P., Bovy, P.H.L., (2004). Pedestrian route-choice and activity scheduling theory and models. Transportation Research Part B-Methodological 38(2), 169-190.

Johansson, A., Batty, M., Hayashi, K., Al Bar, O., Marcozzi, D., Memish, Z.A. (2012). Crowd and environmental management during mass gatherings. *The Lancet infectious diseases*, *12*(2), 150-156.

Johansson, A., Goscè, L. (2014). Utilizing Crowd Insights to Refine Disease-Spreading Models. In *Pedestrian and Evacuation Dynamics 2012* (pp. 1395-1403). Springer, Cham.

Karl, S., Halder, N., Kelso, J.K., Ritchie, S.A., Milne, G.J. (2014). A spatial simulation model for dengue virus infection in urban areas. *BMC infectious diseases*, *14*(1), 447.

Kermack, W.O., McKendrick, A.G. (1927). A contribution to the mathematical theory of epidemics. *Proceedings of the royal society of london. Series A, Containing papers of a mathematical and physical character*, *115*(772), 700-721.

Khan, K., Memish, Z.A., Chabbra, A., Liauw, J., Hu, W., Janes, D.A., ..., Raposo, P. (2010). Global public health implications of a mass gathering in Mecca, Saudi Arabia during the midst of an influenza pandemic. *Journal of travel medicine*, *17*(2), 75-81.

Khan, K., McNabb, S.J., Memish, Z.A., Eckhardt, R., Hu, W., Kossowsky, D., ..., McCloskey, B. (2012). Infectious disease surveillance and modelling across geographic frontiers and scientific specialties. *The Lancet infectious diseases*, *12*(3), 222-230.

Koo, J.R., Cook, A.R., Park, M., Sun, Y., Sun, H., Lim, J.T., Tam, C., Dickens, B.L. (2020). Interventions to mitigate early spread of SARS-CoV-2 in Singapore: a modelling study. *The Lancet Infectious Diseases*.

Kleczkowski, A., Grenfell, B.T. (1999). Mean-field-type equations for spread of epidemics: The 'small world'model. *Physica A: Statistical Mechanics and its Applications*, *274*(1-2), 355-360.

Lewnard, J.A., Lo, N.C. (2020). Scientific and ethical basis for social-distancing interventions against COVID-19. *The Lancet Infectious Diseases*.

Liu, W.M., Hethcote, H. W., Levin, S.A. (1987). Dynamical behavior of epidemiological models with nonlinear incidence rates. *Journal of mathematical biology*, *25*(4), 359-380.

Loh, N.-H.W., Tan, Y., Taculod, J., Gorospe, B., Teope, A.S., Somani, J., Tan, A.Y.H. (2020). The impact of high-flow nasal cannula (HFNC) on coughing distance:

implications on its use during the novel coronavirus disease outbreak. *Canadian Journal of Anesthesi*a, pp. 1-2

Longini, I.M., Nizam, A., Xu, S., Ungchusak, K., Hanshaoworakul, W., Cummings, D.A., Halloran, M. E. (2005). Containing pandemic influenza at the source. *Science*, *309*(5737), 1083-1087.

Milne, G.J., Kelso, J.K., Kelly, H.A., Huband, S. T., McVernon, J. (2008). A small community model for the transmission of infectious diseases: comparison of school closure as an intervention in individual-based models of an influenza pandemic. *PloS one*, *3*(12).

Moussaid, M., Helbing, D., Theraulaz, G. (2011). How simple rules determine pedestrian behavior and crowd disasters. *Proceedings of the National Academy of Sciences of the United States of America*, 108(17), 6884-6888.

Namilae, S., Derjany, P., Mubayi, A., Scotch, M., Srinivasan, A. (2017). Multiscale model for pedestrian and infection dynamics during air travel. *Physical review E*, *95*(5), 052320.

Nuno, M., Chowell, G., Gumel, A.B. (2007). Assessing the role of basic control measures, antivirals and vaccine in curtailing pandemic influenza: scenarios for the US, UK and the Netherlands. *Journal of the Royal Society Interface*, *4*(14), 505-521.

Olsen, S.J., Chang, H.L., Cheung, T.Y.Y., Tang, A.F.Y., Fisk, T.L., Ooi, S.P.L., ..., Hsu, K.H. (2003). Transmission of the severe acute respiratory syndrome on aircraft. *New England Journal of Medicine*, *349*(25), 2416-2422.

Qu, Y.C., Gao, Z.Y., Xiao, Y., Li, X.G. (2014). Modeling the pedestrian's movement and simulating evacuation dynamics on stairs. *Safety Science* 70, 189-201.

Rizzo, C., Lunelli, A., Pugliese, A., Bella, A., Manfredi, P., Tomba, G.S., ..., Degli Atti, M.C. (2008). Scenarios of diffusion and control of an influenza pandemic in Italy. *Epidemiology & Infection*, *136*(12), 1650-1657.

Ross, R. (1910). *The prevention of malaria*. J. Murray.

Schuette, M.C. (2003). A qualitative analysis of a model for the transmission of varicella-zoster virus. *Mathematical biosciences*, *182*(2), 113-126.

World Health Organization (2020). Coronavirus disease 2019 (COVID-19) Situation Report – 72. World Health Organization.

Yang, Y., Atkinson, P.M. (2008). Individual space-time activity-based model: a model for the simulation of airborne infectious-disease transmission by activity-bundle simulation. *Environment and Planning B: Planning and Design*, *35*(1), 80-99.

Yang, Y., Atkinson, P.M., Ettema, D. (2011). Analysis of CDC social control measures using an agent-based simulation of an influenza epidemic in a city. BMC infectious diseases 11(1), 199.

Xiong, C., Hu, S., Yang, M., Younes, H., Luo, W., Ghader, S., Zhang, L. (2020). Data-Driven Modeling Reveals the Impact of Stay-at-Home Orders on Human Mobility during the COVID-19 Pandemic in the US. *arXiv preprint* arXiv:2005.00667

Zhang, L., Ghader, S., Pack, M. L., Xiong, C., Darzi, A., Yang, M., ..., Hu, S. (2020a). An interactive COVID-19 mobility impact and social distancing analysis platform. *medRxiv*.

Zhang, L., Ghader, S., Darzi, A. (2020b). Data Analytics and Modeling Methods for Tracking and Predicting Origin-Destination Travel Trends Based on Mobile Device Data. *Federal Highway Administration Exploratory Advanced Research Program*.

Zheng, Y., Jia, B., Li, X.-G., Jiang, R. (2017). Evacuation dynamics considering pedestrians' movement behavior change with fire and smoke spreading. *Safety Science*, 92, 180-189.

**Appendix A The impact of desired speed**

In this appendix, a series of simulations with different desired speeds are presented in an indoor place with 100 individuals (1 infected) randomly walking inside for 10 minutes. In each simulation, the desired speeds of individuals are respectively set as a fixed value from 0 to 1.6 meter per second. The total infection time $T^{inf}$ (i.e., Eq. (14)) is shown in Figure A.1.

Through the statistics of the individual infections, it is found that the total infection time moves around 700 seconds. It is to say, once the pedestrians are uniformly distributed in the place, the infection time results would be approximately the same and have an insignificant relationship with the desired speed.

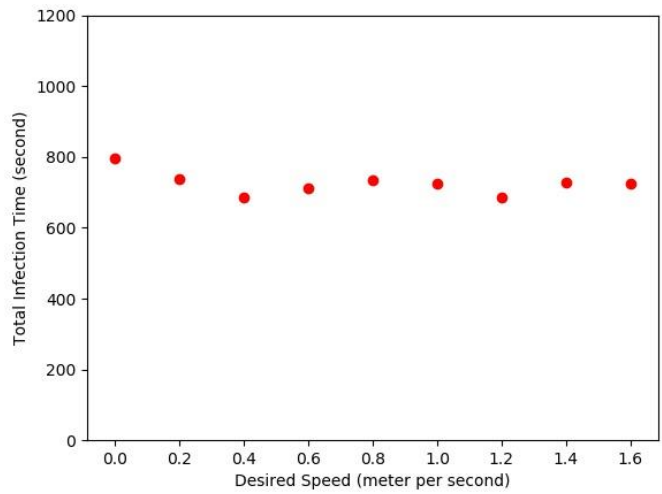

Figure A.1. The total infection time for different desired speeds.